\documentclass[aps,prl,twocolumn,floatfix,superscriptaddress,showpacs,amsfonts,amssymb,amsmath,preprintnumbers]{revtex4}

\usepackage{graphicx}% Include figure files
\usepackage{amssymb}

\newcommand\Alfven{Alfv\'en }

\begin{document}

%\preprint{}

%\title{Reply Comment on ``Kinetic Simulations of Magnetized Turbulence in Astrophysical Plasmas''}
%\begin{center}{ \bf Reply Comment on ``Kinetic Simulations of Magnetized Turbulence in Astrophysical Plasmas''}
%\end{center}


%\maketitle

{\bf Howes {\it et al.} reply:} Matthaeus, Servidio, and Dmitruk
(hereafter MSD) \cite{MSD:2008} comment on our Letter
\cite{Howes:2008a}, arguing that our resolution is insufficient to
claim (i) that our gyrokinetic (GK) simulations capture simultaneously
the MHD and kinetic sub-ion-gyroscale plasma turbulent cascades.  Our
spatial resolution ($64\times 64 \times 128$) is insufficient to
determine converged values of spectral indices---a claim we do not
make---but is sufficient to capture the nonlinear energy transfer from
large to small scales, see a qualitative transition of electric and
magnetic energy spectra at the ion gyroscale, and yield results
consistent with theory above and below this scale (see Figs.~2 and 3
of \cite{Howes:2008a}).  MSD speculate that the low-wave-number
spectrum in our simulations is imposed by the driving.  The driving
occurs at the lowest mode numbers $(k_x,k_y) L_\perp/(2
\pi) = (1,0)$, $(0,1)$, and $(-1,0)$; higher-wave-number 
amplitudes are determined by nonlinear interactions.

Our Letter presents the first kinetic simulation showing that the
spectra of anisotropic turbulence in a weakly collisional plasma are
qualitatively consistent with measurements in the solar wind (SW)
\cite{Bale:2005}, a result not \emph{a priori} obvious
(\cite{Bowers:2007} addresses a different problem).  This first step
in tackling SW turbulence at kinetic scales suggests that GK modeling
is a promising line of inquiry.

MSD allege that we claim (ii) that dissipation is required to achieve
our result.  This apparently arises from a misunderstanding of the
nature of kinetic \Alfven waves (KAW), not a dissipative effect but
dispersive waves weakly Landau-damped at $k_\perp\rho_i>1$. We
emphasize that damping is small for the parameters used in our
simulation (see Fig.~1 of \cite{Howes:2008a}), so the spectrum
attributed to KAW turbulence is near-dissipationless (agreeing with
suggestions in Refs.~\cite{Bale:2005,Alexandrova:2007}, which MSD
incorrectly claim to contradict our results).

Although GK cannot describe all features of SW turbulence (as
we acknowledge \cite{Howes:2008b}), it is a rigorous approximation for
small-scale anisotropic low-frequency fluctuations (see
\cite{Howes:2006} and references therein).  The anisotropy assumption
is based on a wealth of theoretical, numerical, and observational
evidence (see \cite{Howes:2008a}).

While questioning the validity of GK, MSD appear to suggest that Hall
MHD (HMHD) is a preferable model for SW turbulence. Much has been
learned from this temptingly simple model \cite{ref6}.  The fully
nonlinear anisotropic limit ($k_\parallel \ll k_\perp$) of HMHD can be
derived from GK in the limit $\beta_i\ll1$ and $\beta_e \sim
1$ (see Appendix E of Ref.~\cite{Schekochihin:2007}).  The transition
from the MHD limit to KAW occurs at $l_t \sim d_i/\sqrt{1+2/\beta_e}$,
where $d_i=\rho_i/\sqrt{\beta_i}$ is the ion inertial scale.  For $l_t
\gg \rho_i$ (enabling a fluid approximation), we must have
$\beta_i\ll1$ and $T_i\ll T_e$---it is well known from kinetic theory
\cite{ref9} that this {\em cold ion} limit is the only one in which
HMHD is strictly valid, a limit not universally applicable to the SW.
Thus, neither the transition to the dispersive waves nor the effects
of collisionless damping (not captured by any fluid-like limit) are
correctly described by HMHD in the SW parameter regime.

While it is true that the transition to dispersive waves in HMHD
causes flatter electric and steeper magnetic energy spectra, it is
quite unclear whether the simulations MSD present are directly
applicable to the SW (much more unclear, we believe, than in
the case of GK).  Without a kinetic simulation to assess the
importance of finite-ion-gyroscale effects and collisionless damping,
a numerical HMHD simulation may be inconsistent with the
underlying kinetic plasma physics, little more than an
\emph{ad hoc} model.  It is possible to get what appears to be the
right answer for the wrong reason, as demonstrated by the 1-D~HMHD
result shown by MSD: surely they are not claiming that SW turbulence
is accurately modeled as a 1-D cascade in~$k_\parallel$!  That
qualitatively correct spectra seem to emerge in such studies does not
invalidate the much more general GK approach or remove the
need for kinetic simulations.

Obviously, the success of GK in qualitatively reproducing the observed
spectra does not in itself prove its applicability in space.  We make
no such claim and accept that further analytical and numerical work is
needed.  In presenting the first numerical results of our long-term
project on kinetic turbulence, we phrased our conclusions
carefully. The hyperbolic claims attributed to us by
MSD---\emph{e.g.}, that we prove the applicability of GK to the SW, or
show that the spectra reported in Ref.~\cite{Bale:2005} require
dissipation---are an incorrect representation of our work.\\

G.~G.~Howes, S.~C.~Cowley, W.~Dorland, G.~W.~Hammett, E.~Quataert,
A.~A.~Schekochihin and T.~Tatsuno

\end{document}